\documentclass[usenatbib]{mnras}

\usepackage[T1]{fontenc}
\usepackage{ae,aecompl}

\usepackage{graphicx}	

\title[Kinematics and metallicity of SECCO~1]{A very dark stellar system lost in Virgo: kinematics and metallicity of SECCO~1 with MUSE\thanks{Based on data obtained with the European Southern Observatory Very Large Telescope, Paranal, Chile, under the Programme 295.B-5013.
Based on observations made with the NASA/ESA Hubble Space Telescope, obtained at the Space Telescope Science Institute, which is operated by the Association of Universities for Research in Astronomy, Inc., under NASA contract NAS 5-26555. These observations are associated with program GTO-13735.}}

\author[G. Beccari]{
G.Beccari$^{1}$,\thanks{E-mail: gbeccari@eso.org}
M.Bellazzini$^{2}$,
L.Magrini$^{3}$,
L. Coccato$^{1}$,
G. Cresci$^{3}$,
F. Fraternali$^{4}$,
\newauthor
P.T. de Zeeuw$^{1,5}$,
B. Husemann$^{1}$,
R. Ibata$^{6}$, 
G. Battaglia$^{7}$, 
N. Martin$^{6,8}$, 
V. Testa$^{9}$,  
\newauthor
S. Perina$^{2},$
M. Correnti$^{10}$
\\
$^1$European Southern Observatory, Karl-Schwarzschild-Strasse 2, 85748 Garching bei M\"unchen, Germany\\ 
$^2$ INAF - Osservatorio Astronomico di Bologna, Via Ranzani 1, 40127 Bologna, Italy\\
$^3$INAF - Osservatorio Astrofisico di Arcetri, Largo E. Fermi 5, 50125 Firenze, Italy\\
$^4$Dipartimento di Fisica \& Astronomia, Universit\`a degli Studi di Bologna, Viale Berti Pichat, 6/2, I - 40127 Bologna, Italy\\
$^5$Leiden Observatory, Leiden University, Postbus 9513, 2300 RA, Leiden, The Netherlands\\
$^6$Obs. astronomique de Strasbourg, Universit\'e de Strasbourg, CNRS, UMR 7550, 11 rue de l'Universit\'e, F-67000 Strasbourg, France\\ 
$^7$Instituto de Astrofisica de Canarias, 38205 La Laguna, Tenerife, Spain\\             
$^8$Max-Planck-Institut f\"ur Astronomie, K\"onigstuhl 17, D-69117 Heidelberg, Germany\\
$^9$INAF - Osservatorio Astronomico di Roma, via Frascati 33, 00040 Monteporzio, Italy\\ 
$^{10}$Space Telescope Science Institute, Baltimore, MD 21218            
$^{11}$ Universidad de La Laguna, Dpto. Astrofisica, E-38206 La Laguna, Tenerife, Spain\\
}

\date{Accepted 2016 November 4. Received 2016 November 3; in original form 2016 July 6}

\pubyear{2016}

\begin{document}
\label{firstpage}
\pagerange{\pageref{firstpage}--\pageref{lastpage}}
\maketitle

\begin{abstract}
We present the results of VLT-MUSE integral field spectroscopy of SECCO~1, a faint, star-forming stellar system recently discovered as the stellar counterpart of an Ultra
Compact High Velocity Cloud (HVC274.68+74.0), very likely residing within a substructure of the Virgo cluster of galaxies. 
We have obtained the radial velocity of a total of 38 individual compact sources identified as H{\sc ii} regions in the main and secondary body of the system, 
and derived the metallicity for 18 of them. We provide the first direct
demonstration that the two stellar bodies of SECCO~1 are physically associated and that their velocities match the H{\sc i} velocities. The metallicity is
quite uniform over the whole system, with a dispersion 
lower than the uncertainty on individual metallicity estimates. 
The mean abundance, $\langle12+{\rm log(O/H)}\rangle=8.44$, is much higher than the typical values for local dwarf galaxies of similar stellar mass. This strongly suggests 
that the SECCO~1 stars were born from a pre-enriched gas cloud, possibly stripped from a larger galaxy. Using archival HST 
images we derive a total stellar mass of $\simeq 1.6\times 10^5~M_{\sun}$ for SECCO~1, confirming that it has a very high H{\sc i} to stellar mass ratio for a 
dwarf galaxy, M$_{H{\sc i}}$/M$_{*}\sim 100$. The star formation rate, derived from the H$_{\alpha}$ flux is a factor of more than $10$ higher than in typical dwarf galaxies of similar luminosity.
\end{abstract}




\section{Introduction}

SECCO\footnote{\tt http://www.bo.astro.it/secco} \citep[][B15a, hereafter]{pap1} is a survey aimed at searching for stellar counterparts of Ultra Compact High Velocity Clouds (UCHVC) of neutral hydrogen that have been recently proposed by different teams  as candidate mini-halos residing in the Local Group or its surroundings \citep[$D\la 3.0$~Mpc;][ALFALFA and GALFA-H{\sc i} surveys, respectively]{adams,galfa}. The only way to confirm the nature of UCHVCs as small gas-rich dwarf galaxies is to identify a concomitant stellar population that can allow to constrain the distance to the system. Several groups have attempted this search, mainly using public archive data. On the other hand SECCO is based on very-deep, homogeneous, wide-field imaging obtained with the Large Binocular Telescope (MtGraham, AZ), allowing a full quantitative characterisation of non-detections \citep[][B16, hereafter]{pap2}.

In B15a we identified a candidate faint and blue stellar counterpart to an UCHVC from the \citet[][A13 hereafter]{adams} sample (HVC274.68+74.0). The spectroscopic follow-up of the brightest source of the system, by \citet[][B15b, hereafter]{secco1}, showed that it is an H{\sc ii} region at the same velocity\footnote{All the velocity values reported in this letter are heliocentric radial velocities ($V_r$).} of the cloud, $V_r=-128\pm 6$~km/s (A13), hence physically associated to it. As detailed in B15b, having  considered several different hypotheses, we concluded that the newly discovered stellar system, that we baptised SECCO~1, is (most probably) a star-forming and extremely gas-rich ($M_{HI}/L_V\sim 20$) dwarf galaxy located in the Virgo cluster of galaxies. In particular, SECCO~1 likely resides within the substructure named Low Velocity Cloud (LVC) whose members span the range of radial velocity $-400$~km/s$\la V_r\la+400$~km/s \citep[][]{boselli}. 
The coarse metallicity estimate we obtained in B15b was suggestive of
a system much more metal-rich than the typical dwarf of similar luminosity.
\citet{sand15} confirmed our conclusions with independent data but did not provide an independent estimate of the metallicity. On the other hand they identified a second small group of blue stars $\sim 2\arcmin$ Northeast of the main body of SECCO~1, suggesting that the two systems are associated. In the following we will refer to the original stellar system found by B15a as the main body (MB) of SECCO~1, and to the new group as its secondary body (SB). In B16 we showed that SB is indeed recognisable in our original SECCO images and that some additional blue sources possibly associated with SECCO~1 can be identified also to the East of the main body. Moreover, by comparison with very low-surface-brightness dwarfs also residing in Virgo, identified in the same images as SECCO~1, we were able to constrain any population older than $\sim 2$~Gyr within SECCO~1 to have surface brightness significantly fainter than $\mu_V\simeq 27.0$~mag/arcsec$^2$.

Finally, \citet[][A15, hereafter]{adams15} presented new H{\sc i} observations that resolved the original HVC274.68+74.0 cloud (that was seen as an individual entity of size $5\arcmin \times 4\arcmin$ at the $3\arcmin$ resolution of the ALFALFA survey) into a $\simeq 1.75\arcmin \times 1.25\arcmin$ main cloud nearly centred on MB (AGC~226067), likely connected by a thin bridge to a smaller cloud ($\simeq 0.75\arcmin \times 0.75\arcmin$, AGC~229490) located $\simeq 0.5\arcmin$ to the West of SB. A15 concluded that (a) with the revised estimates of the mass of the gas clouds the gas-to-stellar-luminosity(mass) ratio becomes less extreme than that derived by B15b and S15, and more similar to the galaxies studied by \citet{can_shield}, and (b) that SECCO~1 MB (AGC~226067), SB (which they call AGC 229491) and the nearby cloudlet AGC~229490 are two or three low-mass galaxies within the Virgo cluster having some kind of mutual interaction \citep[see A15 for a detailed discussion, and][for a theoretical analysis]{bekki}.

Here we present the velocity field and the metallicity of several H{\sc ii} regions in SECCO~1 MB and SB obtained with the Multi Unit Spectroscopic Explorer (MUSE) at the VLT. Archival images data taken with the Hubble Space Telescope (HST) are used in support of our study. A thorough analysis of the full wealth of information contained in the MUSE and HST data is deferred to a future contribution, where we will discuss in depth the various hypotheses for the origin of SECCO~1. 
According to Mei et al. (2007) the Virgo cluster has a significant depth along the line of sight 
(the $\pm3\sigma$ range is $\sim4$ Mpc). Hence it should be kept in mind that the distance of SECCO~1 is not so tightly constrained by its 
membership to Virgo; the LVC probably lies slightly behind the core of the cluster~\citep[][]{mei}. In the following we adopt the conventional 
distance D=17.0 Mpc, after~\citet[][]{boselli} and consistent with A15.


\section{Observations and data reduction}
\label{sec:obs}
   \begin{figure*}
   \centering
   \includegraphics[width=0.8\textwidth]{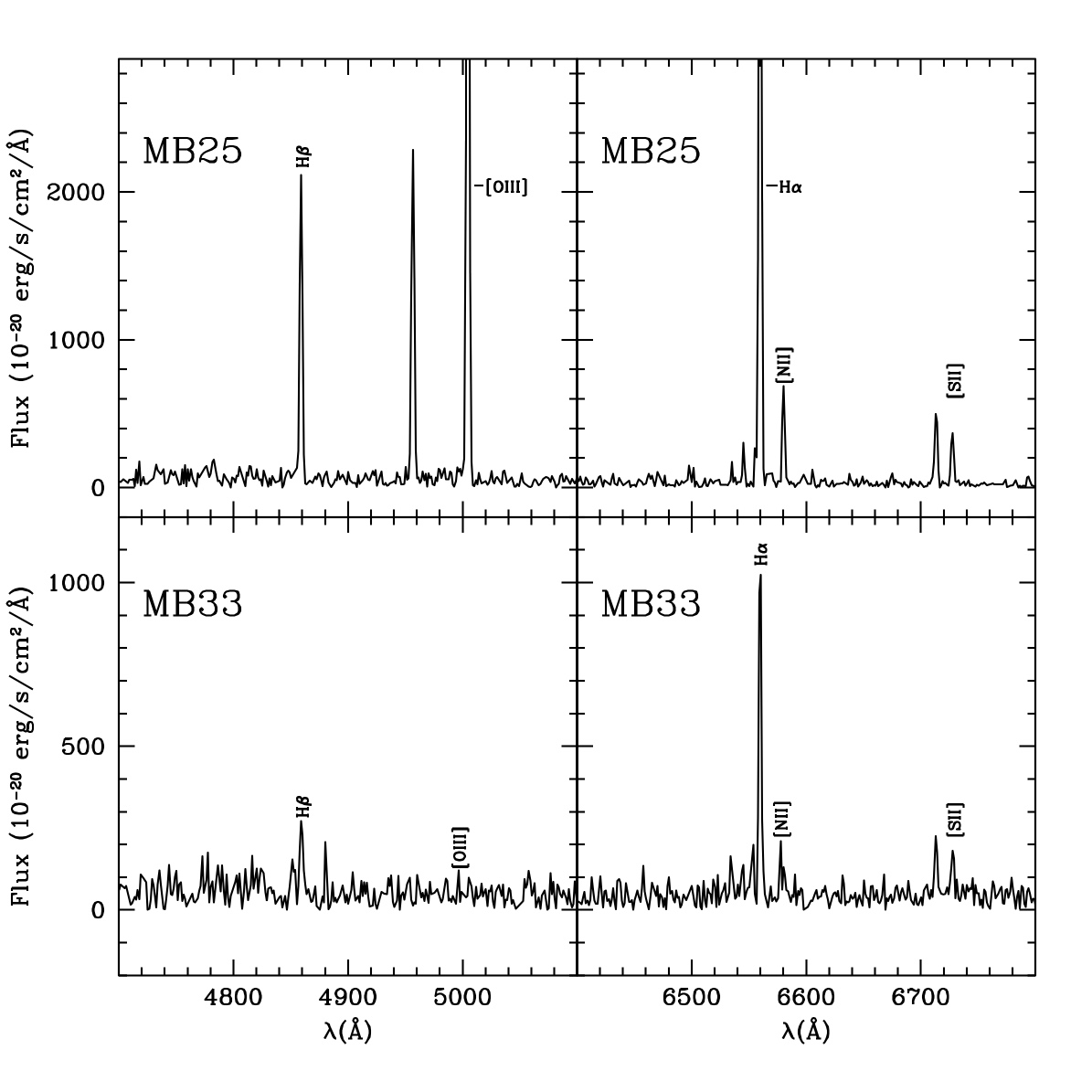}
     \caption{Portions of the extracted spectra of two compact sources identified in the MUSE data, namely MB25 and MB33, 
     having the highest and lowest S/N respectively.  
     The name of the source is indicated in the plots and follows the nomenclature adopted in in Table~\ref{tabVr}. Some of emission lines are labeled.}
        \label{fig_spec}
    \end{figure*}

MUSE is a panoramic integral-field spectrograph able to acquire low resolution spectra ($R=2000-4000$, from the bluest to the reddest wavelength) in the spectral range $0.465-0.93\mu$m~\citep[][]{bac14}.
The adopted Wide Field Mode (WFM) provides a field of view (FoV) of $1\arcmin\times1\arcmin$, with a pixel scale $0.2\arcsec/$pix. A total of 4~h of telescope time were allocated  to the Director's  Discretionary Time
program 295.B-5013 (PI: Beccari).

Two pointings were sufficient to sample the MB and the SB of SECCO~1.  For each target we 
acquired 6 exposures of 1000~s each. We used a simple dither pattern and applied 90 deg offset to the de-rotator between each consecutive exposure, to improve the flat fielding and to ensure an homogeneous image quality across the FoV. The average seeing during the
exposures was $0.7\arcsec$, corresponding to a measured full width at half maximum (FWHM) of $\sim 3.5$~px on the final image.
The raw data from each exposure were reduced and combined with the MUSE pipeline v1.4~\citep[][]{wel12} run under the esoreflex environment~\citep[][]{fre13}. The final cubes were split into 3801 single layers (i.e. single frames), sampling the targets from $4600.29\AA$ to $9350.29\AA$ with a
step in wavelength of $1.25\AA$. 
 
The modest degree of crowding in our data allowed us to extract the background subtracted 
flux of any source from every single layer of the cube using a standard aperture photometry routine with the task {\tt phot} of {\sc iraf} . 
A master list of objects was created by searching for individual sources with significant peaks ($> 5\sigma$ above the background) a) in an image obtained  collapsing few frames located around the H$\alpha$ line, and also b) in an image obtained by collapsing the entire cube into a single frame, and then merging together the two lists. This approach allowed us to recover either sources dominated by H$\alpha$ emission or by the stellar continuum.
The extraction of the spectra was performed using an aperture of radius=5 pix, i.e. $\sim1.5\times$FWHM. 
A larger aperture of 10~pix was also adopted for flux calibration purposes. With this method we extracted the spectra of 68 and 45 sources in the MB and SB, respectively. Further analysis revealed that only 26(12) of them can be reliably associated to MB(SB), the remaining ones being background galaxies or sources lacking spectral features allowing a reliable velocity estimate. These 26+12 =38 bona-fide SECCO~1 sources, listed in Table~\ref{tabVr}, have spectra typical of H{\sc ii} regions and are the subject of the following analysis.
It cannot be excluded that some of the listed sources are not independent individual H{\sc ii} regions but are in fact emission peaks within larger complexes. For the purpose of the present analysis this does not seem to be a serious concern. A complementary view not based on integrated spectra of individual sources but on a pixel-by-pixel analysis will be presented in a forthcoming paper (Magrini et al. in prep).

In Fig.~\ref{fig_spec} we show two illustrative examples of the 1-D spectra extracted from the MUSE cubes, bracketing the range of signal-to-noise ratio (S/N) where we were able to derive reliable metallicity estimates (see Sect.~\ref{met}, below). The superb sensitivity of MUSE, the possibility to integrate the entire flux of each source inherent to 2-D spectrographs, and the lack of any slit-aligmnent issue (that plagued B15b observations) makes the quality and the S/N of the MUSE spectra presented here much higher than those analysed in B15b (see their figure 2 for comparison). Moreover the spectral range covered by MUSE spectra is much wider than in B15b, e.g., including H$_{\beta}$, that was not reached in B15b spectra. All these factors imply a significant improvement in the precision and reliability of the new metallicity estimates for sources of similar luminosity, with respect to that analysis (see Sect.~\ref{met}, for further discussion).

   \begin{figure*}
   \centering
   \includegraphics[width=12truecm]{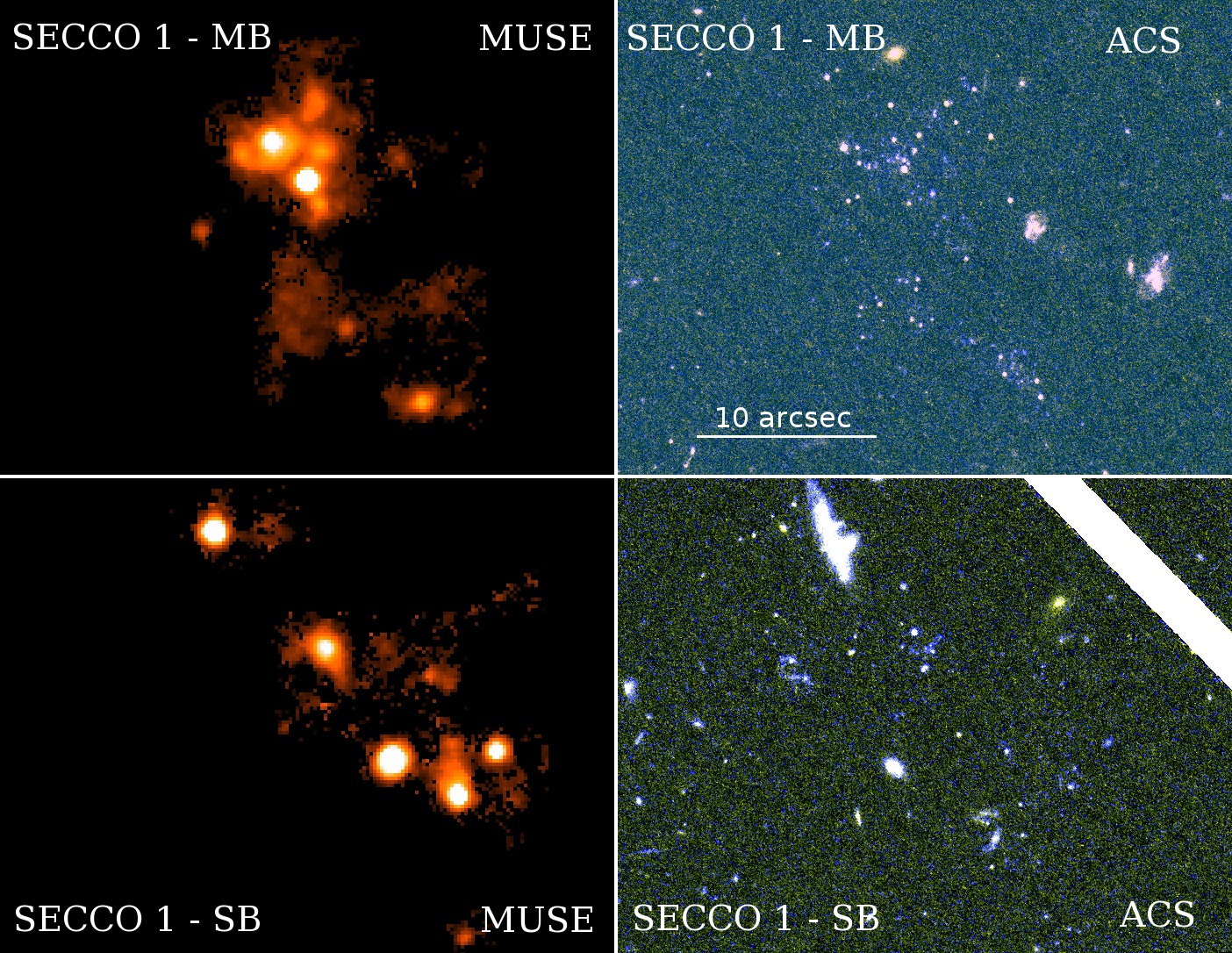}
     \caption{Left panels: Continuum-subtracted $H\alpha$ images 
     centered on the main and secondary body of SECCO~1 (upper and lower panel, respectively). Right panels: Color image of the same field from the F606W and F814W HST images. The diagonal white band in the lower right panel is the inter-chip gap of the camera. North is up, East to the left.}
        \label{ima}
    \end{figure*}

The left panels of Fig.~\ref{ima} show stamp-size H$\alpha$ images of SECCO~1 MB and SB obtained from the MUSE cube. It is clear that, in addition to several individual sources (likely H{\sc ii} regions) there is also diffuse hot gas emitting in H$\alpha$. In the right panels, we show the corresponding portions of the combined F606W ($t_{\rm exp}=2196$~s) and F814W ($t_{\rm exp}=2336$~s) images from the HST GO-13735 program (PI: Sand), that we retrieved from the STScI 
archive. 
These images resolve many of the structures seen in the H$\alpha$ images into individual bright and very blue compact sources. The correspondence between the sources identified in the MUSE image and those seen in the ACS ones is excellent.
We note that four SB sources are superposed on the prominent background disk galaxy at the upper margin of the image in the lower right panel of Fig.~\ref{ima}. The lines of this galaxy (at z=0.187), as well as those of the more distant smaller spiral the lower end  of its disc (at z=0.752), are both seen in the spectra of these SB sources, a remarkable case of multiple chance superposition.
No underlying over-density of old stars (i.e. Red Giant Branch stars, with age$\ga 2$~Gyr) associated with SECCO~1 is readily visible in the ACS image, in agreement with the conclusions by B16.

%
\begin{table*}
	\centering
	\caption{Position and velocity for SECCO1 HII regions}
	\label{tabVr}
	\begin{tabular}{lccccc} 
		\hline
   Name   &  RA$_{J2000}$ & Dec$_{J2000}$&  V$_r^{a}$      & N$_{l}^{b}$& H$_{\alpha}$ Flux$^{c}$ \\
          &	[deg]	  &	[deg]	 &  [km/s]     &	 & 10$^{-18}$ erg cm$^{-2}$ s$^{-1}$   \\
\hline
   MB14   &   185.47292   & 13.45734     &$-171\pm 7  $& 2  &$ 29.7\pm	4.5   $       \\
   MB15   &   185.47360   & 13.45784     &$-176\pm 11 $& 3  &$ 54.9\pm  5.7   $       \\
   MB25   &   185.47537   & 13.46103     &$-154\pm 4  $& 7  &$319.3\pm 19.0   $       \\
   MB26   &   185.47644   & 13.46143     &$-152\pm 8  $& 3  &$171.6\pm 11.6   $       \\
   MB27   &   185.47583   & 13.46145     &$-152\pm 5  $& 7  &$296.9\pm 17.8   $       \\
   MB28   &   185.47683   & 13.46176     &$-151\pm 6  $& 5  &$ 61.3\pm  6.1   $       \\
   MB29   &   185.47600   & 13.46184     &$-152\pm 6  $& 7  &$218.9\pm 13.9   $       \\
   MB30   &   185.47559   & 13.46189     &$-152\pm 5  $& 6  &$226.4\pm 14.3   $       \\
   MB31   &   185.47440   & 13.46191     &$-154\pm 4  $& 2  &$ 33.3\pm  4.7   $       \\
   MB32   &   185.47605   & 13.46241     &$-153\pm 9  $& 3  &$ 43.4\pm  5.2   $       \\
   MB33   &   185.47515   & 13.46244     &$-151\pm 8  $& 4  &$ 78.2\pm  6.9   $       \\
   MB34   &   185.47470   & 13.46269     &$-151\pm 30 $& 1  &$ 26.9\pm  4.3   $       \\
   MB38   &   185.47599   & 13.46325     &$-160\pm 30 $& 1  &$ 15.3\pm  3.8   $       \\
   MB39   &   185.48480   & 13.46324     &$-167\pm 30 $& 1  &$  9.1\pm  2.5   $       \\
   MB56   &   185.47591   & 13.46160     &$-153\pm 4  $& 7  &$259.3\pm 16.0   $       \\
   MB57   &   185.47710   & 13.46022     &$-159\pm 30 $& 1  &$ 11.2\pm  3.6   $       \\
   MB58   &   185.47473   & 13.45869     &$-169\pm 30 $& 1  &$ 35.9\pm  4.8   $       \\
   MB59   &   185.47516   & 13.46063     &$-157\pm 6  $& 7  &$233.3\pm 14.7   $       \\
   MB60   &   185.47386   & 13.46140     &$-162\pm 30 $& 1  &$ 23.2\pm  4.2   $       \\
   MB61   &   185.47513   & 13.46149     &$-152\pm 5  $& 7  &$266.6\pm 16.3   $       \\
   MB62   &   185.47534   & 13.46228     &$-143\pm 7  $& 5  &$104.6\pm  8.2   $       \\
   MB63   &   185.47349   & 13.45754     &$-156\pm 10 $& 3  &$ 64.7\pm  6.2   $       \\
   MB64   &   185.47299   & 13.45942     &$-211\pm 15 $& 3  &$ 26.0\pm  4.3   $       \\
   MB65   &   185.47556   & 13.45894     &$-143\pm 11 $& 2  &$ 59.3\pm  6.0   $       \\
   MB66   &   185.47582   & 13.45961     &$-151\pm 30 $& 1  &$ 33.4\pm  4.7   $       \\
   MB68   &   185.47523   & 13.46229     &$-148\pm 7  $& 5  &$ 97.9\pm  7.9   $       \\
          &	          &	         &	       &    & 	      	   	      \\       
   SB01   &   185.48451   & 13.48588     &$-153\pm 6  $& 6  &$170.7\pm 11.5   $       \\
   SB02   &   185.48465   & 13.48644     &$-151\pm 7  $& 6  &$185.3\pm 12.3   $       \\
   SB03   &   185.48285   & 13.48419     &$-134\pm 11 $& 4  &$109.5\pm  8.5   $       \\
   SB33   &   185.48427   & 13.48577     &$-152\pm 12 $& 3  &$113.9\pm  8.7   $       \\
   SB34   &   185.48272   & 13.48436     &$-126\pm 11 $& 4  &$102.7\pm  8.1   $       \\
   SB35   &   185.48314   & 13.48445     &$-121\pm 12 $& 4  &$ 90.6\pm  7.5   $       \\
   SB36   &   185.48105   & 13.48392     &$-126\pm 15 $& 3  &$ 54.1\pm  5.7   $       \\
   SB38   &   185.47995   & 13.48273     &$-116\pm 11 $& 3  &$ 99.5\pm  8.0   $       \\
   SB39   &   185.48163   & 13.48259     &$-123\pm 9  $& 5  &$169.9\pm 11.5   $       \\
   SB41   &   185.48058   & 13.48202     &$-133\pm 8  $& 5  &$154.8\pm 10.7   $       \\
   SB42   &   185.48046   & 13.47975     &$-134\pm 30 $& 1  &$ 13.2\pm  3.7   $       \\
   SB45   &   185.48455   & 13.48620     &$-150\pm 6  $& 6  &$193.3\pm 12.7   $       \\
\hline
\multicolumn{5}{l}{$^a$ Heliocentric radial velocity.}\\
\multicolumn{5}{l}{$^b$ Number of emission lines used to estimate the radial velocity.}\\
\multicolumn{5}{l}{$^c$ Observed flux, not extinction corrected.}\\
	\end{tabular}
\end{table*}

\section{Radial velocity}
\label{rv}

We derived the radial velocity of each individual compact source
by determining the wavelength shift of the available emission lines relative to their rest
wavelengths. To estimate the peak of each emission line we used  a gaussian fitting algorithm.
The final radial velocity was obtained from the average wavelength shift; the velocity uncertainties, given by the RMS divided by the square root of the number of measured lines, encompass the range 
$\simeq 2-30$~km/s, with an average value of $\simeq 10$~km/s (see Table~\ref{tabVr}).

\begin{table*}
  \begin{center}
  \caption{Mean Properties of stars and neutral hydrogen in SECCO~1}
  \label{mean}
  \begin{tabular}{lcccccc}
\hline
name &  $\langle V_r\rangle$& $\sigma$ & note & $\langle 12+{\rm log}(O/H)\rangle^a$ & note& $V_{int}^b$\\
     &   [km/s]          & [km/s] &   &  & & [mag]\\
\hline
SECCO~1 MB   & $-153.2\pm 1.4^c$    & $3.5\pm 2.1^c$ &from 25 H{\sc ii} regions$^e$  & $8.37\pm 0.11$  & from 4 H{\sc ii} regions$^d$ & $20.9\pm 0.4$\\
AGC226067& $-142$     & $9 \pm 3$ &H{\sc i} (from A15)  &   &  &\\
\hline	      
SECCO~1 SB   & $-126.5\pm 2.5^c$    & $2.7\pm 5^c$ &from 8 H{\sc ii} region$^e$  &  $8.39\pm 0.11$  & from 5 H{\sc ii} regions$^d$ &$22.0\pm 0.5$\\
AGC229490& $-123$     &$4 \pm 2$  &H{\sc i} (from A15)  &   &  &\\
\hline	      
\multicolumn{7}{l}{$^a$ Straight mean $\pm$ standard deviation}\\
\multicolumn{7}{l}{$^b$ Integrated apparent magnitude from aperture photometry on the ACS images}\\
\multicolumn{7}{l}{$^c$ Computed with a Maximum Likelihood algorithm that take into account the effect of errors on individual velocities}\\
\multicolumn{7}{l}{$^d$ Only the sources where a reliable metallicity estimate can be obtained and not affected by contamination from nearby regions
, see Tab.~\ref{tabMet}}\\
\multicolumn{7}{l}{$^e$ MB: Excluding MB64. SB: Excluding the four most northern sources, that are superposed to two background galaxies}\\
\end{tabular} 
\end{center}
\end{table*}

In Table~\ref{mean} we report the mean velocity and metallicity of SECCO~1 MB and SB, separately, and the velocities of the associated H{\sc i} clouds, from A15, for comparison. The reported mean velocities and velocity dispersions have been computed with a Maximum Likelihood (ML) algorithm that finds the most likely parameters of a gaussian model taking into account (and correcting for) the effects of the errors on individual velocities, following \citet{mart07}.
The small difference in mean velocity \citep[$27\pm3$~km/s, to be compared with the velocity dispersion of the LVC, $\sigma_{LVC}=208$~km/s,][]{boselli} implies that MB and SB are indeed physically associated. MB and the AGV226067 H{\sc i} cloud have compatible systemic radial motions; the same is true for SB and AGC229490, in spite of the spatial offset between the two systems. On the other hand, we cannot exclude a small ($\la 10$~km/s) but real mismatch between the systemic velocity of the MB and SB H{\sc ii} regions and the corresponding H{\sc i} clouds, as the small spatial offset between SB and AGC229490 may suggest ongoing ejection of the neutral gas from their stellar counterparts.
The velocity dispersion of the two stellar systems is only marginally resolved in our data. This may be partly due to the contamination among spectra of overlapping sources, especially in the densest region of MB, that can result in spurious correlations of velocities. Moreover, we have excluded from the ML analysis one source from the MB sample and four sources from the SB sample, having discrepant velocities (see below). 
If we compute the unweighted mean and standard deviation from all the measured sources we obtain $\langle V_r\rangle=-157.7$~km/s and $\sigma=13.5$~km/s for MB, and $\langle V_r\rangle=-134.9$~km/s and $\sigma=13.3$~km/s for SB, leaving substantially unchanged the above conclusions and indicating that
the true dispersion should be similar to that measured in H{\sc i}. The isolated source at (X,Y)$_{\rm kpc}\simeq (-2.5,0.5)$ provides significant support to the hypothesis that a few additional components associated with SECCO~1 are indeed present to the East of MB, as suggested by B16.

\begin{figure*}
   \centering
   \includegraphics[width=0.8\textwidth]{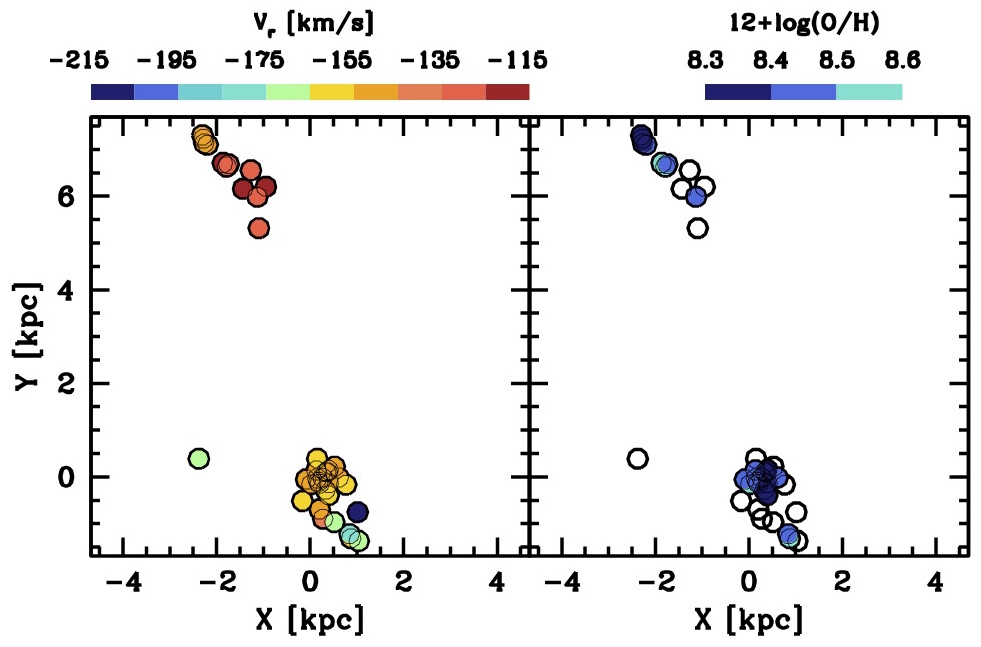}
     \caption{Left panel: Color-coded velocity map for the 38 SECCO~1 MB and SB sources identified in our MUSE data. The origin of the X,Y axes is taken as the center of MB as derived in B15b.
     Right panel: the same as above but with the sources color-coded according to their metallicity.
     Empty circles are sources whose metallicity cannot be reliably estimated from their spectra.
     Note that the range of metallicity variation has an amplitude similar to the uncertainty on individual 12+log(O/H) estimates.}
        \label{mapvel}
    \end{figure*}

\begin{figure}\centering
   \includegraphics[width=0.49\textwidth]{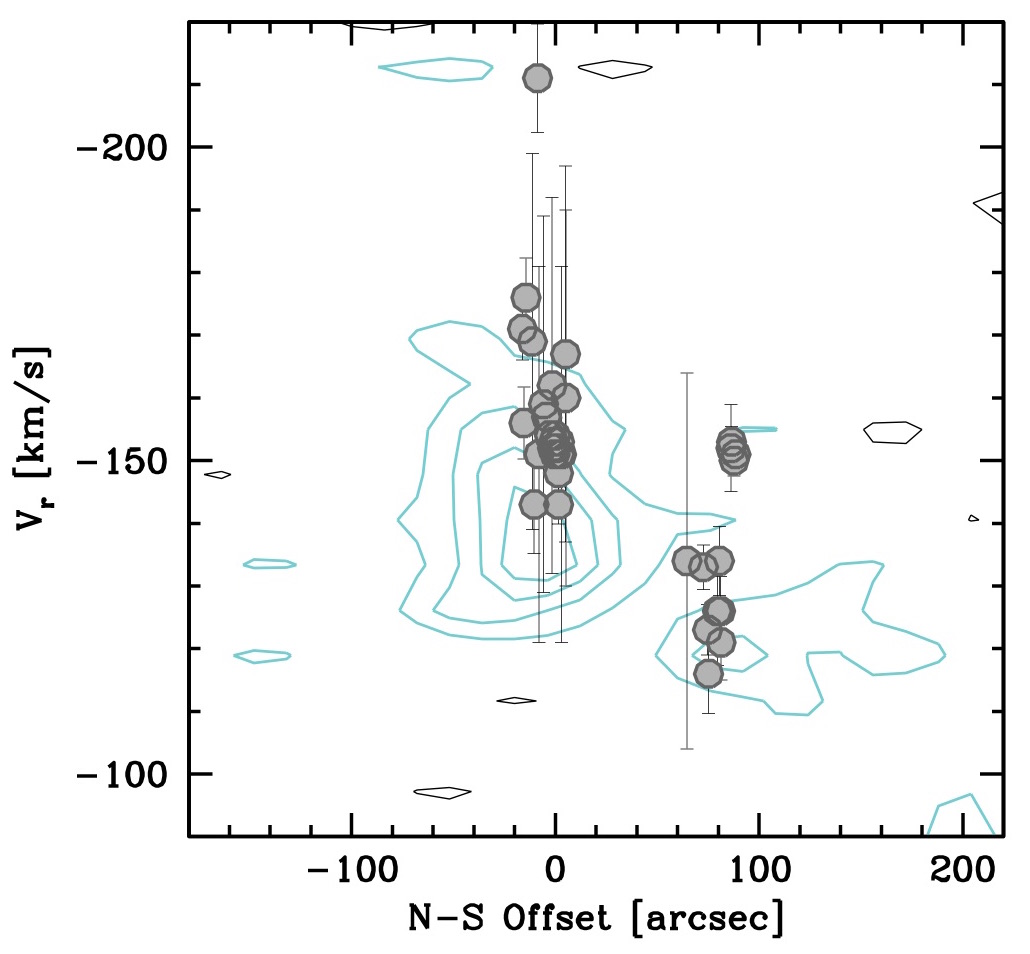}
     \caption{A position-velocity slice along the North-South direction of the H{\sc i} data by A15 is compared to the velocity of the H{\sc ii} regions identified in this paper (grey filled circles). The H{\sc i} intensity contours are plotted at -2 (black), 2, 4, 6, 8 (green) mJy~bm$^{-1}$ .}
        \label{velH2}
    \end{figure}

The left panel of Fig.~\ref{mapvel} shows that the H{\sc ii} regions within MB display a weak velocity gradient very similar to that displayed by the H{\sc i} (A15). In general, the overall velocity fields traced by the H{\sc ii} regions and by the neutral hydrogen are remarkably similar, within the uncertainties. A direct comparison is presented in Fig.~\ref{velH2}, where our sources are superimposed on the North-South position-velocity slice we derived from the A15 data (kindly provided by E. A. K. Adams). A few sources do not follow the general pattern, including, e.g. the source with $V_r<-205$~km/s at (X,Y)$_{\rm kpc}\simeq (1,-1)$, and the four SB sources superposed on a background spiral at (X,Y)$_{\rm kpc}\simeq (-2.5,7)$. It is interesting to note that these velocity outliers seem to have H{\sc i} counterparts, albeit of weak intensity.
To check the significance of these sources we used ${\rm 3D}$Barolo~\citep[][]{di15} to run a source detection in the HI data-cube.
The only two sources detected at high resolution with reliable sensitivities (thresholds between 2.5 and 3.0 $\sigma_{\rm r.m.s.}$ were Secco 1 MB 
and a cloud a much smaller velocity ($v_{\rm r}=-65.6 {\rm km}\,{\rm s}^{-1}$ at RA=185.466  an DEC=13.459) probably belonging to the Milky Way 
environment, if real. At lower spatial resolution ($>60"$) more sources are found, some emission appears in the region of the SB and further out with 
some hint of a general gradient in velocity. However it is very hard with these data to assess its reliability.

%
\begin{table}
	\centering
	\caption{Line flux in SECCO1 HII regions in units of the H$_{\beta}$ flux, set to H$_{\beta}$=100.}
	\label{tabFlux}
	\begin{tabular}{lccccc} 
		\hline

Name &[OIII]          &H$_{\alpha}$  &[NII]         &[SII]        &[SII]       \\
     &5007\AA         &6563\AA       &6584\AA       &6717\AA      &6730\AA     \\ 
		\hline
MB14 &-           		    &289$\pm$20    &-        &-        &-   \\
MB15 &73$\pm$25       &289$\pm$40    &32$\pm$23     &84$\pm$27    &33$\pm$23   \\
MB25 &256$\pm$16      &289$\pm$18    &32$\pm$4      &33$\pm$4	  &25$\pm$4    \\
MB26 &56$\pm$9        &289$\pm$23    &52$\pm$9      &56$\pm$9	  &43$\pm$9    \\
MB27 &130$\pm$10      &289$\pm$20    &44$\pm$6      &50$\pm$6	  &37$\pm$5    \\
MB28 &68$\pm$34       &289$\pm$60    &32$\pm$33     &71$\pm$38    &34$\pm$34   \\
MB29 &60$\pm$7        &289$\pm$20    &54$\pm$7      &55$\pm$7	  &39$\pm$6    \\
MB30 &118$\pm$11      &289$\pm$22    &52$\pm$8      &51$\pm$8	  &39$\pm$7    \\
MB31 &125$\pm$38      &289$\pm$48    &52$\pm$34     &59$\pm$35    &43$\pm$34   \\
MB32 &51$\pm$22       &281$\pm$33$^a$    &35$\pm$21     &85$\pm$23    &49$\pm$22   \\
MB33 &29$\pm$17       &289$\pm$35    &19$\pm$17     &57$\pm$19    &49$\pm$19   \\
MB56 &75$\pm$7        &289$\pm$19    &52$\pm$6      &54$\pm$7	  &38$\pm$6    \\
MB59 &238$\pm$16      &289$\pm$18    &28$\pm$5      &32$\pm$5	  &26$\pm$5    \\
MB61 &282$\pm$18      &289$\pm$18    &41$\pm$5      &37$\pm$5	  &29$\pm$5    \\
MB62 &69$\pm$15       &289$\pm$30    &58$\pm$15     &59$\pm$15    &48$\pm$15   \\
MB63 &31$\pm$16       &289$\pm$30    &39$\pm$17     &80$\pm$19    &41$\pm$17   \\
MB68 &70$\pm$16       &289$\pm$33    &64$\pm$17     &58$\pm$17    &48$\pm$16   \\
SB01 &112$\pm$12      &289$\pm$22    &29$\pm$7      &23$\pm$7	  &19$\pm$7    \\
SB02 &99$\pm$10       &289$\pm$19    &27$\pm$6      &27$\pm$6	  &21$\pm$6    \\
SB03 &15$\pm$9        &289$\pm$24    &47$\pm$11     &60$\pm$12    &38$\pm$10   \\
SB33 &136$\pm$16      &289$\pm$27    &41$\pm$12     &19$\pm$11    &24$\pm$11   \\
SB34 &17$\pm$9        &265$\pm$20$^a$    &26$\pm$9      &-  	  & -	       \\
SB35 &31$\pm$12       &289$\pm$25    &42$\pm$12     &55$\pm$13    &39$\pm$12   \\
SB41 &80$\pm$11       &289$\pm$24    &41$\pm$9      &31$\pm$9	  &28$\pm$8    \\
SB45 &100$\pm$10      &289$\pm$20    &25$\pm$6      &26$\pm$6	  &22$\pm$6    \\
\hline
\multicolumn{5}{l}{Only lines used for metallicity estimates are included.}\\
\multicolumn{5}{l}{$^a$ Negative extinction constant, null extinction assumed.}\\
	\end{tabular}
\end{table}

\section{Metallicity}
\label{met}

All measured line intensities were corrected for extinction computing the ratio between the observed and theoretical Balmer decrement for the typical conditions of an H{\sc ii} region 
\citep[see][]{os06}. In the few cases, where the extinction constant is negative we assumed a null extinction. 
The extinction in both the MB and SB is quite low, with average c$\beta$ in the MB and SB equal to 0.29$\pm$0.24 and 0.22$\pm$0.26, respectively. 
We measured the line fluxes with the task {\tt splot} of {\sc iraf}. 
In the MUSE spectral range, we measured 
recombination lines of H (H$\alpha$ and H$\beta$) and collisional lines of several ions ([O{\sc iii}], [N{\sc ii}], [S{\sc ii}], 
and in a few cases [O{\sc ii}], [Ar{\sc iii}], and [S{\sc iii}]). All the diagnostic plots based on the ratios between available lines consistently classify all the identified sources as H{\sc ii} regions. 
The fluxes of the lines that were used for the estimate of the metallicity  are presented in Table~\ref{tabFlux}, in a scale where the flux of $H_{\beta}$ is conventionally set to 100. The conversion to physical units can be obtained with the $H_{\alpha}$ fluxes provided in Table~\ref{tabVr}, as some sources lack a reliable estimate of the $H_{\beta}$ flux.

Due to the absence of electron-temperature diagnostic lines, the gas-phase oxygen abundance of each source is determined 
with three different {\em strong-line} ratios: 
N2=[N{\sc ii}]/H$\alpha$, O3N2=([O{\sc iii}]/H$\beta$)/([N{\sc ii}]/H$\alpha$) 
\citep[][hereafter PP04]{pp04} and O3S2N2, a combination of the line ratios R3=([O{\sc iii}]($\lambda$4959+$\lambda$5007))/H$\beta$, N2, and NS=([S{\sc ii}]($\lambda$6717+$\lambda$6730))/H$\alpha$, also known as S2~\citep[][]{pm11}.
N2 and O3N2 from PP04 have indeed proved to be 
the strong-line diagnostics that give metallicity values very close to those obtained by directly measuring 
the electron temperature of the gas through [O~{\sc iii}]4363 line \citep[see, e.g.][]{am13}.
On the other hand, when [O{\sc iii}], [S{\sc ii}], [N{\sc ii}], H$\alpha$, and H$\beta$ are all simultaneously available with good signal-to-noise ratios, we take advantage of the NS method that allows us to estimate the ionising field in absence of the  [O {\sc ii}] lines and has been successfully validated against oxygen abundances computed using the electron temperature~\citep[][]{lo12}. Finally, we compute the average of the abundances from N2 and O3N2. We combine N2 and O3N2 to compensate for the effect of a varying ionisation.
We rejected all metallicity estimates including line ratios with S/N$<3.0$. Once this selection is applied we get reliable estimates of the metallicity, at least from one indicator, for 18 sources, 10 in MB and 8 in SB.

All the metallicity estimates from individual indicators as well as the adopted mean from different indicators are reported in
Table~\ref{tabMet}, where we list also the estimates obtained from N2 and O3N2 (and their average) adopting the calibration by \citet{m13}, instead of \citet{pp04}. Table~\ref{tabMet} provides all the information to evaluate the (negligible) effect of adopting one of the different indicators and/or calibrations chosen on our results. 
Six sources whose line ratios was found to change varying from 5 px to 10 px the radius of the aperture used of the flux's extraction are marked with and asterisk in 
Table~\ref{tabMet}, and have been excluded when computing the average metallicity obtained from estimates involving the [OIII] line. In fact, the changes are small 
and limited to line ratios including [OIII] and are due to contamination from the two very bright adjacent sources MB25 and MB61, that have very strong [OIII] lines. 
We verified, using the~\citet[][]{dop16} diagnostic diagrams, that the higher EW([OIII]) in these sources is not associated to metallicity effects but to an increase of 
the ionisation parameter. A detailed discussion of the ionisation conditions across SECCO 1 will be presented in a forthcoming paper (Magrini et al. in prep.).
The last four rows of the table report the straight mean (Mean) and standard deviation ($\sigma$), the mean and standard deviation obtained with with the same ML algorithm used in Sect.~\ref{rv} (Mean$_{ML}$ and $\sigma_{ML}$) with the associated uncertainties, and the number of sources used to derive these parameters (N).

Note that each calibration has an intrinsic uncertainty of about $\sim$0.2 dex and that they are not exactly on the same absolute scale; it is reassuring that, nevertheless, they provide fully consistent results. 
In particular, the 
spread in metallicity is always consistent with zero, within the uncertainties: even the straight standard deviation is smaller than the uncertainty on the individual 12+log(O/H) estimates, independently of the adopted indicator, implying a strong degree of chemical homogeneity among the various SECCO~1 sources.
In the following we will adopt metallicities from the combination of the PP04's methods as our reference values (second column of Tab.~\ref{tabMet}). 
The mean abundance from this combination of indicators is ${\langle \rm 12+log(O/H)\rangle}=8.38$, corresponding to $0.5Z_{\sun}$ \citep[assuming the solar abundance of][]{grev}. The average values for MB and SB are the same within the uncertainty, ${\langle \rm 12+log(O/H)\rangle}=8.37\pm 0.11$ and ${\langle \rm 12+log(O/H)\rangle}=8.39\pm 0.11$, respectively (see Tab.~\ref{mean}).

The mean abundance of SECCO~1  is much higher than in dwarf galaxies with  similar stellar mass \citep[see, e.g.,][and B15b]{lee03}, being more typical of galaxies as conspicuous as as M33~\citep{ma10} or the LMC~\citep{ca08,pagel}. 
It is also interesting to note that the observed abundance of SECCO~1 H{\sc ii} regions is significantly higher than the hot intra-cluster medium at the same projected distance from the center of Virgo \citep[$\simeq 0.1Z_{\sun}$,][]{urban}.

In the right panel of Fig.~\ref{mapvel} the sources for which a reliable estimate of the metallicity can be obtained are color-coded according to their 12+log(O/H) values. The map illustrates very clearly the 
chemical homogeneity of all the sources, already noted above, with no detectable trend with position or velocity. 

%

%
\begin{table*}
	\centering
	\caption{Metallicity of SECCO1 HII regions from different indicators and calibrations}
	\label{tabMet}
	\begin{tabular}{lccccccc} 
		\hline
Name          &$\langle$12+log(O/H)$\rangle^a$	  &12+log(O/H)  &12+log(O/H)		 & 12+log(O/H)    &12+log(O/H)      &12+log(O/H)  & $\langle$12+log(O/H)$\rangle^b$    \\
Ind(Cal)      &N2+O3N2(PP04)   & N2(PP04)	&O3N2(PP04)		 & NS(PM11)	  & N2(M13)	    & O3N2(M13)       & N2+O3N2(M13)   \\
              &		  &		&			 &		  &		    &		      & 	       \\
MB25	      &8.33 $\pm$ 0.21 &  8.36 $\pm$ 0.08 &  8.30  $\pm$0.11 & 8.31 $\pm$  0.18 & 8.30  $\pm$  0.08  & 8.25 $\pm$ 0.11 &  8.28  $\pm$  0.19  \\
MB26$^*$      &8.52 $\pm$ 0.22 &  8.48 $\pm$ 0.10 &  8.57  $\pm$0.17 & 8.33 $\pm$  0.22 & 8.40  $\pm$  0.10  & 8.43 $\pm$ 0.17 &  8.41  $\pm$  0.20  \\
MB27$^*$      &8.43 $\pm$ 0.21 &  8.44 $\pm$ 0.09 &  8.43  $\pm$0.11 & 8.33 $\pm$  0.17 & 8.37  $\pm$  0.09  & 8.34 $\pm$ 0.11 &  8.35  $\pm$  0.19  \\
MB29$^*$      &8.53 $\pm$ 0.21 &  8.49 $\pm$ 0.08 &  8.57  $\pm$0.14 & 8.35 $\pm$  0.18 & 8.41  $\pm$  0.08  & 8.42 $\pm$ 0.14 &  8.42  $\pm$  0.19  \\
MB30$^*$      &8.47 $\pm$ 0.22 &  8.48 $\pm$ 0.10 &  8.47  $\pm$0.13 & 8.37 $\pm$  0.20 & 8.40  $\pm$  0.10  & 8.36 $\pm$ 0.13 &  8.38  $\pm$  0.19  \\
MB56	      &8.50 $\pm$ 0.21 &  8.47 $\pm$ 0.09 &  8.53  $\pm$0.12 & 8.35 $\pm$  0.17 & 8.40  $\pm$  0.09  & 8.40 $\pm$ 0.12 &  8.40  $\pm$  0.19  \\
MB59	      &8.30 $\pm$ 0.22 &  8.32 $\pm$ 0.10 &  8.28  $\pm$0.13 & 8.26 $\pm$  0.22 & 8.27  $\pm$  0.10  & 8.24 $\pm$ 0.13 &  8.26  $\pm$  0.20  \\
MB61	      &8.36 $\pm$ 0.21 &  8.42 $\pm$ 0.08 &  8.31  $\pm$0.11 & 8.36 $\pm$  0.18 & 8.35  $\pm$  0.08  & 8.26 $\pm$ 0.11 &  8.30  $\pm$  0.19  \\
MB62$^*$	      &8.53 $\pm$ 0.25 &  8.50 $\pm$ 0.15 &  8.56  $\pm$0.25 & 8.36 $\pm$  0.33 & 8.42  $\pm$  0.15  & 8.42 $\pm$ 0.25 &  8.42  $\pm$  0.23  \\
MB68$^*$	      &8.54 $\pm$ 0.26 &  8.52 $\pm$ 0.16 &  8.57  $\pm$0.26 & 8.46 $\pm$  0.42 & 8.44  $\pm$  0.16  & 8.42 $\pm$ 0.26 &  8.43  $\pm$  0.23  \\
SB01	      &8.37 $\pm$ 0.24 &  8.34 $\pm$ 0.14 &  8.40  $\pm$0.18 & 8.29 $\pm$  0.35 & 8.29  $\pm$  0.14  & 8.31 $\pm$ 0.18 &  8.30  $\pm$  0.22  \\
SB02	      &8.36 $\pm$ 0.23 &  8.31 $\pm$ 0.12 &  8.40  $\pm$0.17 & 8.24 $\pm$  0.28 & 8.27  $\pm$  0.12  & 8.31 $\pm$ 0.17 &  8.29  $\pm$  0.21  \\
SB03	      &9999 $\pm$  999 &  8.45 $\pm$ 0.13 &  9999  $\pm$ 999 & 8.25 $\pm$  0.28 & 8.38  $\pm$  0.13  & 9999 $\pm$  999 &  9999  $\pm$   999  \\
SB33	      &8.41 $\pm$ 0.26 &  8.41 $\pm$ 0.16 &  8.41  $\pm$0.22 & 8.29 $\pm$  0.49 & 8.35  $\pm$  0.16  & 8.32 $\pm$ 0.22 &  8.34  $\pm$  0.23  \\
SB34	      &9999 $\pm$  999 &  8.33 $\pm$ 0.18 &  9999  $\pm$ 999 & 9999 $\pm$   999 & 8.28  $\pm$  0.18  & 9999 $\pm$  999 &  9999  $\pm$   999  \\
SB35	      &9999 $\pm$  999 &  8.43 $\pm$ 0.16 &  9999  $\pm$ 999 & 8.25 $\pm$  0.33 & 8.36  $\pm$  0.16  & 9999 $\pm$  999 &  9999  $\pm$   999  \\
SB41	      &8.45 $\pm$ 0.24 &  8.42 $\pm$ 0.13 &  8.49  $\pm$0.19 & 8.34 $\pm$  0.31 & 8.35  $\pm$  0.13  & 8.37 $\pm$ 0.19 &  8.36  $\pm$  0.21  \\
SB45	      &8.35 $\pm$ 0.24 &  8.30 $\pm$ 0.13 &  8.39  $\pm$0.18 & 8.23 $\pm$  0.30 & 8.26  $\pm$  0.13  & 8.31 $\pm$ 0.18 &  8.28  $\pm$  0.21  \\
\hline
Mean          &8.38            &  8.41            &  8.39            & 8.32             & 8.35       	     & 8.31            &  8.31               \\
$\sigma$      &0.06            &  0.07            &  0.08            & 0.06             & 0.06         0.05  & 0.04            &  0.05    	      \\
Mean$_{ML}$   &8.38 $\pm$ 0.07 &  8.42 $\pm$ 0.02 &  8.37  $\pm$0.05 & 8.32 $\pm$  0.06 & 8.35  $\pm$  0.03  & 8.30 $\pm$ 0.05 &  8.31  $\pm$  0.07  \\
$\sigma_{ML}$ &0.00 $\pm$ 0.08 &  0.00 $\pm$ 0.03 &  0.00  $\pm$0.07 & 0.00 $\pm$  0.06 & 0.00  $\pm$  0.03  & 0.00 $\pm$ 0.05 &  0.00  $\pm$  0.07  \\
    N         & 9   	       &  18              &   9    	     &   17             &  18   	     &  9              &   9                  \\
\hline
\multicolumn{8}{l}{PP04=\citet{pp04}; PM11=\citet{pm11}; M13=\citet{m13}. The NS indicator is also known as S2.}\\
\multicolumn{8}{l}{$^a$ Average of the values from N2(PP04) and O3N2(PP04). Adopted as reference in the analysis.}\\
\multicolumn{8}{l}{$^b$ Average of the values from N2(M13) and O3N2(M13). Reported for comparison.}\\
\multicolumn{8}{l}{$^*$ [OIII] flux possibly affected by contamination from nearby bright sources in r=10~px apertures.}\\
	\end{tabular}
\end{table*}

\section{Stellar Mass and Star Formation Rate}
\label{phot}
The very irregular morphology of SECCO1 as well as the lack of detection of an underlying unresolved population makes the estimate of the integrated luminosity quite challenging. Our MUSE data give us the unprecedented capability of distinguishing sources that are members of SECCO~1 from unrelated background galaxies over the whole extension of the system. This factor, coupled with the use of deep high-resolution ACS images, allow us to obtain a new estimate of the total luminosity of SECCO~1 that is much more robust and reliable than previous ones (e.g., B15b, A15).
We used the Aperture Photometry Tool \citep[APT,][]{apt} to estimate the total (sky-subtracted) flux within a few non-overlapping circular apertures of different radii ($\sim 5\arcsec-10\arcsec$) properly placed to enclose all the MB and SB sources and their surroundings while avoiding contamination by nearby background sources (mainly distant galaxies). The sky was estimated as the median value over wide annuli surrounding each aperture. The resulting F606W and F814W integrated magnitudes were transformed into V magnitudes using Eq.~1 of \citet{silvia}. We obtain $V_{int}=20.9 \pm 0.5$ and $V_{int}=22.0 \pm 0.5$ for MB and SB, respectively. The F606W-F814W colors are $\simeq 0.1\pm 0.5$ and 
$\simeq 0.0\pm 0.5$, respectively. Adopting E(B-V)=0.048 from B15a and D=$17.0\pm 1.0$~Mpc, the absolute magnitudes are $M_V=-10.4\pm 0.4$ and $M_V=-9.3\pm 0.5$ for MB and SB, respectively, corresponding to $L_V=1.2^{+0.5}_{-0.4}\times10^{6}~L_{V,\sun}$ and  $L_V=4.4^{+2.6}_{-1.6}\times10^{5}~L_{V,\sun}$; the total V luminosity of SECCO1 (MB+SB) is $L_V=1.6^{+0.6}_{-0.4}\times10^{6}~L_{V,\sun}$. The new $M_V$ value for MB is one full magnitude fainter than what obtained by B15b and 0.7 mag fainter than that reported by A15\footnote{Converted from M$_g$ and $(g-i)_0$ with the equation $V=g-0.390(g-i)-0.032$, derived in the same way as Eq.~5 and Eq.~6 by \citet{vv124}, and valid for $-0.6\le g-i\le 2.0$.}, with the same assumptions on distance and reddening. The difference is due to the removal of the contribution of background sources from the luminosity budget of SECCO1, which was not possible with the data available to B15a or A15. Note however, that, because of the considerable uncertainty associated with these measures, the newly derived value is within $\le 2\sigma$ of previous estimates.

From the above numbers and the new H{\sc i} mass estimates by A15, considering only the spatially coincident AGC226067 H{\sc i} cloud and assuming a 10 per cent error in $M_{HI}$, we obtain, 
for the main body of SECCO~1 $\frac{M_{HI}}{L_V}=12.5^{+6.5}_{-4.5}$. A similar ratio is obtained considering SB and AGC 229490, but given the spatial offset between the two systems it is unclear how meaningful $\frac{M_{HI}}{L_V}$ is in this specific case. 

The fact that all the detected sources are H{\sc ii} regions implies that the stellar population largely dominating the light from SECCO~1 has an age $\la 30$~Myr. There is general consensus that such a population should have a stellar $M/L_V$ ratio $\la 0.1$, with $M/L_V$ decreasing for younger ages. For example, the models by \citet[][]{maraston} predict $M/L_V=0.09513$ for a Simple Stellar Population (SSP) of age=30~Myr with [Z/H]=-1.35, and $M/L_V=0.09170$ for [Z/H]=-0.33, assuming a  \citet{salp} Initial Mass Function (IMF). The corresponding numbers for a \citet{kroupa} IMF are $M/L_V=0.06205$ and $M/L_V=0.05957$. The BASTI solar-scaled models \citep{basti1,basti2} predicts $M/L_V=0.061(0.083)$ for SPSS with age=30(50)~Myr and [Fe/H]=-0.66, and $M/L_V=0.067(0.085)$ for [Fe/H]=-1.49. $M/L_V\simeq 0.06(0.09)$ for SPSS with age=30(50)~Myr is also predicted by the widely used \citet{bc03} models, with negligible dependence on metallicity and on the adopted set of stellar models (see, e.g., their Fig.~1 and Fig.~2). Assuming, conservatively and for simplicity, $M/L_V=0.1$, we obtain a stellar mass $M_{\star}\simeq 1.2\times 10^5~M_{\sun}$ for MB and $M_{\star}\simeq 1.6\times 10^5~M_{\sun}$ for the whole SECCO~1 system.

Given the very low stellar mass, the number of H{\sc ii} regions found in SECCO~1 may appear quite large, if taken at face value (but see Sect.~\ref{sec:obs}, above). For example, the relatively large dIrr WLM has a stellar mass $M_{\star}=4.3\times 10^7~M_{\sun}$ and a H{\sc i} mass $M_{HI}=6.1\times 10^7~M_{\sun}$ \citep{mcc} and it hosts ``only'' 21 H{\sc ii} regions \citep{hm95}. We verified that the observed number of H{\sc ii} regions in SECCO~1 does not imply an anomalous IMF. Adopting, conservatively, an IMF as bottom-heavy as the \citet{salp}, a population with the stellar mass of $M_{\star}\simeq 1.6\times 10^5~M_{\sun}$ is expected to produce $\simeq 800$ stars with $m\ge 10~M_{\sun}$, more than sufficient to populate all the observed H{\sc ii} regions. 

On the other hand, integrating the extinction-corrected $H_{\alpha}$ luminosity over all the 38 H{\sc ii} regions and adopting the calibration by \citet{kenn}, we obtain a total star formation rate (SFR) of $7.2\times10^{-2}~M_{\sun}~{\rm yr}^{-1}$\footnote{The much lower value reported in B15a was derived from the only two H{\sc ii} regions considered in that paper, hence it was based on a severely incomplete census of the $H_{\alpha}$ luminosity of the system as a whole.}. This is significantly higher than the typical SFR observed in dwarf galaxies of comparable luminosity \citep[$\simeq 1.0\times10^{-3}~M_{\sun}~{\rm yr}^{-1}$;][see, e.g., their Fig.~6]{bdd}, suggesting that the ongoing star formation episode that made SECCO~1 detectable is indeed exceptionally strong for such a low-mass system.

\section{Discussion and Conclusions}
\label{conc}

From the results described above we can draw several important conclusions shedding new light onto the nature of SECCO~1. 

First of all, it is demonstrated beyond any doubt that MB, SB (and possibly also a few additional and nearly-isolated H{\sc ii} regions in the surroundings) are physically associated. The remarkable homogeneity in metallicity strongly suggest that the various pieces of SECCO~1 originated from a single, chemically homogeneous H{\sc i} cloud that is now disrupting, probably because of its interaction with the cluster environment. This would make the system unlikely to be in dynamical equilibrium, precluding a dynamical mass estimate

Second, the new observations seem to confirm, albeit with the large associated uncertainties, that the system has an extreme under-abundance of stars given its H{\sc i} content, in agreement with B15b and S15, and contrary to the conclusions of A15. The stellar mass estimate obtained in Sect.~\ref{phot}
implies that the H{\sc i}-to-stellar mass ratio is $\frac{M_{HI}}{M_{\star}}\sim 100$, placing SECCO~1 straight in the realm of almost-dark galaxies, as defined by \citet[][see also \citealt{eckert}]{can_dark}. 

Finally, the very high mean metallicity given the stellar mass ($M_{\star}\sim 1.2\times 10^5~M_{\sun}$, for MB, and $M_{\star}\sim0.4\times 10^5~M_{\sun}$, for SB) strongly suggests that the gas from which SECCO~1 stars were born was pre-enriched elsewhere, possibly torn apart by tides from the disk of a spiral galaxy during its in-fall into Virgo. Indeed, the location of SECCO~1 in a luminosity-metallicity plot is typical of tidal galaxies \citep{sweet}. The (apparent) lack of an old population would also be consistent with this hypothesis. The extreme SFR, given the total luminosity, may be suggestive of a star formation episode induced by some kind of interaction.
As noted in B15a there are no obvious candidate parent galaxies in the immediate surroundings of SECCO~1. An intriguing candidate, for the kind of system that can have produced SECCO~1 as a tidal galaxy is the interacting pair NGC~4299 + NGC~4294 \citep{chung}, which show prominent H{\sc i} tails, have velocity and distance compatible with membership to the LVC, and lie at the projected distance of $\simeq 600$~kpc from SECCO~1. However, an origin from this interacting system would place the detachment of SECCO~1 about 1.2~Gyr in the past (assuming the current radial velocity difference as a reference value), implying a long traveling within Virgo/LVC before the occurrence of the first feeble burst of star formation, just a few tens of Myr ago. This simple consideration suggests that other evolutionary paths are worth to be considered for this intriguing object, that may hint at the existence of an additional class of extremely dark stellar systems. A thorough comparative discussion of alternative hypothesis on the origin of SECCO~1 is clearly beyond the scope of the present contribution and is deferred to a future paper.


\section*{Acknowledgements}

We thank the anonymous referee for her/his very thorough reading of our manuscript, and suggestions that substantially 
improved our paper. We warmly thank Elizabeth Adams for kindly providing the VLA H{\sc i} data-cube of AGC~226067.
G.B. gratefully acknowledges the financial support by the Spanish Ministry of Economy 
and Competitiveness under the Ram\'on y Cajal Program (RYC-2012-11537).
This research has made use of the SIMBAD database, operated at CDS, Strasbourg and 
of the NASA/IPAC Extragalactic Database (NED) which is operated by the Jet Propulsion Laboratory, California Institute of Technology, under contract with NASA.







\bsp	
\label{lastpage}
\end{document}